\documentclass[twocolumn, twocolappendix, trackchanges]{aastex7}

\usepackage{booktabs}
\usepackage{threeparttable} 
\usepackage{amsmath}

\newcommand{\xmm}{XMM-Newton }

\newcommand{\blue}{\textcolor{blue}}

\defcitealias{Pan2024ApJS}{X-LEAP I}
\defcitealias{Qu2024ApJ}{X-LEAP II}

\begin{document}
	
\title{The XMM-Newton Line Emission Analysis Program (X-LEAP) III: Earth's Magnetospheric X-ray Emission Revealed by 22-Year XMM-Newton Observations}
\correspondingauthor{Zhijie Qu}
\email{quzhijie@tsinghua.edu.cn}
	
\author[0000-0002-9943-1155]{Zeyang Pan}
\affiliation{Key Laboratory of Optical Astronomy, National Astronomical Observatories, Chinese Academy of Sciences, 20A Datun Road, Beijing, 100101, People's Republic of China}
\affiliation{School of Astronomy and Space Science, University of Chinese Academy of Sciences, Beijing, 100049, People's Republic of China}
\email{} 


\author[0000-0002-2941-646X]{Zhijie Qu}
\affiliation{Department of Astronomy, Tsinghua University, Beijing 100084, People’s Republic of China}
\email{}

\author{Yuqi Gong}
\affiliation{State Key Laboratory of Space Weather, National Space Science Center, Chinese Academy of Sciences, Beijing 100190, People's Republic of China}
\email{}

\author{Tianran Sun}
\affiliation{State Key Laboratory of Space Weather, National Space Science Center, Chinese Academy of Sciences, Beijing 100190, People's Republic of China}
\email{}

\author[0000-0001-6276-9526]{Joel N. Bregman}
\affiliation{Department of Astronomy, University of Michigan, Ann Arbor, MI 48109, USA}
\email{jbregman@umich.edu}

\author{Yingjie Zhang}
\affiliation{State Key Laboratory of Space Weather, National Space Science Center, Chinese Academy of Sciences, Beijing 100190, People's Republic of China}
\email{}

\author{Li Ji}
\affiliation{Purple Mountain Observatory, Chinese Academy of Sciences, 10 Yuanhua Road, Nanjing 210023, People's Republic of China}
\email{}

\author{Jifeng Liu}
\affiliation{Key Laboratory of Optical Astronomy, National Astronomical Observatories, Chinese Academy of Sciences, 20A Datun Road, Beijing, 100101, People's Republic of China}
\affiliation{School of Astronomy and Space Science, University of Chinese Academy of Sciences, Beijing, 100049, People's Republic of China}
\affiliation{New Cornerstone Science Laboratory, National Astronomical Observatories, Chinese Academy of Sciences, Beijing, 100012, People's Republic of China}
\email{}

\begin{abstract}

The magnetosphere, protecting the Earth from intense solar activity, is also shaped by the solar wind, while its structure is still uncertain in observation.
In this study, we map the X-ray emission in the magnetosphere, which is induced by the charge exchange between the highly-ionized solar wind and the neutral gas around the Earth, known as the magnetospheric solar wind charge exchange (SWCX).
In particular, we extract the magnetospheric SWCX in the \ion{O}{7} line emission data adopted from the \xmm Line Emission Analysis Program (X-LEAP). 
The observed magnetospheric SWCX shows an enhanced emission of approximately $I_{\rm OVII}^{\rm mag}\approx 2$ photons $\rm cm^{-2}~ s^{-1}~sr^{-1}$ toward the Sun, showing a consistent shape predicted by numerical simulations.
Furthermore, this magnetospheric SWCX exhibits a dependence on the \xmm pointing direction, which traces the path length of SWCX emission in the Earth's magnetosphere.
Building on this directional dependence, we model the 3D magnetosheath structure using soft X-ray observations for the first time, constraining the averaged boundary geometry and SWCX emissivity distribution over 22 years.
Finally, utilizing the \xmm data, we derive an empirical \ion{O}{7} emission efficiency of $\alpha_{\rm OVII}=(2.1\pm0.4) \times 10^{-16}\ {\rm eV\,cm^{2}}$.

\end{abstract}
	


\section{Introduction} 
\label{Intro}

The dynamic interaction between the solar wind (SW) and the Earth's magnetosphere is fundamental to understanding the space weather, as the coupling mediates a continuous exchange of energy and momentum between SW and the magnetosphere \citep[see the reviews by][]{Sibeck2018, kuntz2019solar}.
In particular, the SW shapes the magnetosphere by compressing its dayside and elongating its nightside, while the magnetosphere in turn decelerates and deflects the incident SW plasma, thereby regulating the local neutral gas environment \citep[e.g.,][]{Borovsky2018SGeo...39..817B}.
The precise mechanisms of this bidirectional interaction, however, remain poorly quantified due to observational and modeling limitations.

Solar wind charge exchange (SWCX) imaging provides a promising new diagnostic for probing the dynamic interaction between the SW and the magnetosphere \citep[e.g.,][]{Cravens2001JGR, Sibeck2018, Collier2018JGRA..12310189C, SMILE2025SSRv..221....9W}.
SWCX occurs when highly charged SW ions capture electrons from neutrals and emit soft X-rays during de-excitation \citep[e.g.,][]{Cravens1997, Koutroumpa2009}. The resulting emission scales with the neutral gas density ($n_{\rm H}$), the ion density ($n_{\rm SW}$), and the collision speed ($v_{\rm col}$) as $I \propto n_{\rm H}n_{\rm SW}v_{\rm col}$ \citep[e.g.,][]{Cravens1997, zhang2022ApJ...932L...1Z, zhang2023ApJ...948...69}.
Previous studies revealed enhanced X-ray emissions when the line of sight (LOS) traverses the dayside magnetosheath and cusp, i.e., regions known for strong SW-neutral interactions \citep[e.g.,][]{snowden2004ApJ...610.1182S, fujimoto2007evidence, Carter2011A&A, Kuntz2015ApJ}. 
These findings confirm the existence and spatial variability of magnetospheric SWCX, but its global distribution and underlying structure remain poorly understood.

The cumulative \xmm archive over two decades provides a unique opportunity to advance observations on the Earth's magnetosphere.
In particular, the highly elliptical orbit and large effective area of \xmm are crucial for the necessary magnetospheric coverage and efficient detection of diffuse X-ray signals \citep{Jansen2001A&A}. 
Two decades of observations have yielded an extensive archive suited for statistical analysis. 

Utilizing this archive, we established the \xmm Line Emission Analysis Program (X-LEAP) to study the SWCX and the diffuse Milky Way (MW) hot gas emission, by measuring \ion{O}{7}, \ion{O}{8}, and Fe-L line intensities in the soft X-ray background (\citealt{Pan2024ApJS}; hereafter \citetalias{Pan2024ApJS}).
However, there exists significant degeneracy among different emission components, making effective decomposition a crucial issue in the analysis of diffuse soft X-ray backgrounds. In \citetalias{Pan2024ApJS}, we investigated the solar-cycle variation of heliospheric SWCX, while \citet[][hereafter, \citetalias{Qu2024ApJ}]{Qu2024ApJ} focused on the influence of temperature variations in the MW hot gas on the observed X‑ray emission.
In this study, we explore the spatial distribution of magnetospheric SWCX using the X-LEAP data, as another major component in the diffuse X-ray emission.

\begin{figure*}[t!]
\centering
\includegraphics[width=0.90\textwidth]{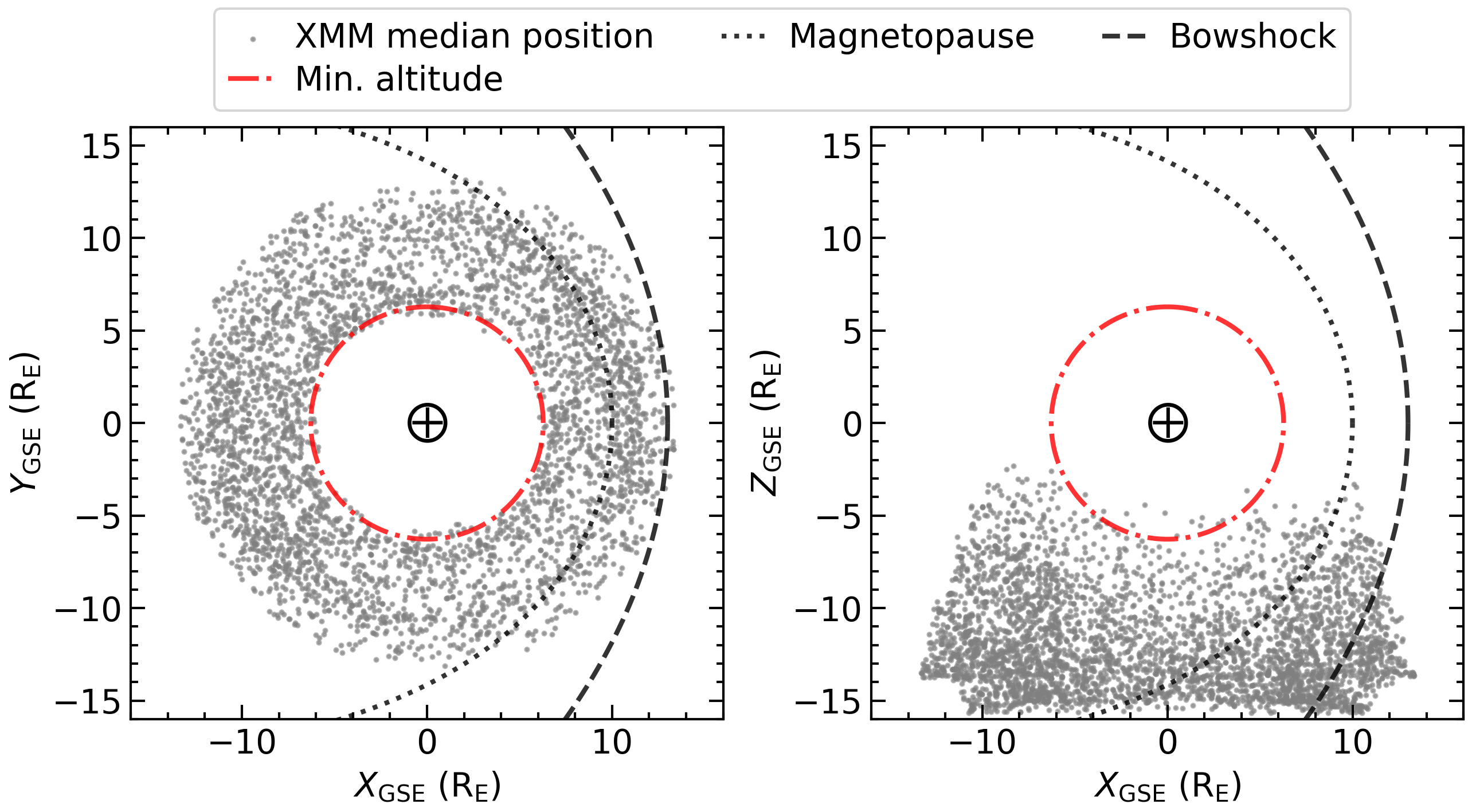}
\caption{The distribution of \xmm observation positions (points) used in this study. These observations are distributed across the magnetosphere, well-suited for probing magnetospheric SWCX. The magnetosheath region, where strong SWCX emission is expected, is bounded by the modeled magnetopause (dotted line; \citealt{Shue1997}) and bow shock (dashed line; \citealt{Chao2002COSPA..12..127C}). The dash-dotted curve indicates XMM-Newton's minimum scientific altitude.}
\label{fig1:data distribution}
\end{figure*}

\section{DATA} \label{sec:data}
In this study, we leverage both the vast \xmm archival data and the latest numerical simulations to constrain the 22-year averaged magnetospheric SWCX.
The \xmm data are adopted from \citetalias{Pan2024ApJS}, which focuses on stable solar activity periods by removing soft proton flares using the XMM-SAS\footnote[1]{https://www.cosmos.esa.int/web/xmm-newton/sas}  \texttt{espfilt} task (This filtering process preserves 76\% of the total exposure time).
This study thereby provides a static reference for future dynamic studies during strong solar flares.

\subsection{\xmm Data} \label{subsec:observation data}

We adopt the “clean” \ion{O}{7} subsample optimized for diffuse background studies from the X-LEAP survey \citepalias{Pan2024ApJS}. 
This subsample is composed of 3,723 EPIC-MOS observations satisfying four key criteria: (1) good time intervals $>$5 ks after filtering soft proton flares; (2) effective regions $>$50\% of the FOV, where masked regions includes (i) bright point sources identified by the XMM-SAS \texttt{cheese} task and (ii) areas overlapping with the halo (i.e., $r_{500}$) of known galaxy clusters or nearby galaxies listed in the MCXC \citep{piffaretti2011mcxc} and the \citet{kourkchi2017galaxy} galaxy catalogs; (3) at Galactic latitudes $|b| > 2^\circ$, minimizing contamination from bright and complex emission in the Galactic plane; and (4) out of large-scale X-ray structures, including the eROSITA bubbles \citep{predehl2020detection}, the Cygnus and Eridanus superbubbles \citep{cash1980x, ochsendorf2015nested}, and the Vela and Monogem supernova remnants \citep{helfand2001vela, thorsett2003pulsar}.

For each observation, the \ion{O}{7} emission is modeled as a Gaussian line together with foreground and background X-ray emission, and the full spectral model is introduced in \citetalias{Pan2024ApJS}. 
The best-fit parameters are optimized under a Bayesian framework
using a Markov Chain Monte Carlo (MCMC) method.
The \ion{O}{7} line intensity $I_{\rm OVII}$ is derived from the normalization of this Gaussian line. 
Normally, the median and $1\sigma$ uncertainty can be extracted from the MCMC chain after the MCMC thermalization.
Here, we use the last 2,000 points in the posterior chain to represent the full probability distribution function in the following analysis.

The derived $I_{\rm OVII}$ ranges from $\sim$2 to 12 photons $\rm cm^{-2}~ s^{-1}~sr^{-1}$ (line units, LU; 5th–95th percentiles), with a median intensity of 5.4 LU and a median $1\sigma$ measurement uncertainty of 0.7 LU. This intensity includes contributions from the magnetospheric SWCX, heliospheric SWCX, and the MW hot gas. 
The latter two components are modeled and subtracted based on their distinct properties, detailed in Sections~\ref{app:helio} and ~\ref{app:MW}. 
After removing their contributions from the posterior chain for every single \xmm observation, the residuals are dominated by the magnetospheric SWCX, so denoted as $I^{\rm mag}_{\rm OVII}$.


The analysis of the magnetosphere requires knowing the satellite's LOS geometry (i.e., its position and pointing direction), since the measured intensity is integrated emission along the viewing direction. 
For consistent identification of magnetospheric SWCX regions, we perform our analysis in the Geocentric Solar Ecliptic (GSE) coordinate system. 
In this system, the $X$-axis points directly from Earth to the Sun, the $Y$-axis lies in the ecliptic plane pointing toward dusk, and the positive $Z$-axis points toward the north ecliptic pole. 

For each observation, the satellite position is represented as the median of its GSE trajectory using the XMM-SAS \texttt{orbit} task. These median positions form a ring-like distribution in the $X_{\rm GSE}$--$Y_{\rm GSE}$ plane, with radii ranging from $\approx$6 to 13 $\rm R_E$ (Figure~\ref{fig1:data distribution}). 
The inner boundary corresponds to XMM-Newton's minimum science altitude, while the outer limit reflects its orbit configuration.
Furthermore, all positions lie at $Z_{\rm GSE}<0$, as the apogee is located below the GSE equatorial plane.
These positions span the region inside the magnetopause, the dayside magnetosheath, and the region outside the magnetosphere, making them well-suited for probing magnetospheric SWCX.

The GSE pointing direction is derived from the observation timestamp and celestial coordinates (RA, Dec) defined in the ICRS frame in two steps. First, the GSE axes at a given epoch are determined within the ICRS frame using the Sun's geocentric position and its apparent geocentric velocity vector (i.e., the time derivative of the position) to construct the orthogonal rotation matrix $R(t)$ \citep{Hapgood1992P&SS...40..711H}. Then, the observed (RA, Dec) is projected into an ICRS unit vector $\hat{n}_\mathrm{ICRS}$, and rotated through $\hat{n}_\mathrm{GSE} = R(t)\hat{n}_\mathrm{ICRS}$ to yield the GSE pointing direction. This transformation is performed using the astropy Python library \citep{Astropy2013A} via \texttt{SkyCoord.transform\_to(GeocentricSolarEcliptic)}, which handles the rotation internally.

XMM-Newton's pointing is constrained to directions within $\pm20^{\circ}$ of the perpendicular to the Sun-Earth line to ensure sufficient energy supply and thermal stability while avoiding direct solar exposure \citep[e.g.,][]{Schartel2022hxga.book..114S}. 
Consequently, all our derived directions satisfy this operational constraint.
The pointing is also constrained to $>42.5^\circ$ from the Earth to avoid limb contamination, resulting in none of the derived sightlines passing within $\sim$8 $\rm R_{\rm E}$ of Earth. Therefore, they are unlikely to capture X-ray emissions from the polar cusps, typically located within this radius \citep[e.g.,][]{fujimoto2007evidence, Sun2019JGRA}.

\begin{figure*}[t!]
	\centering
	\includegraphics[width=0.465\textwidth]{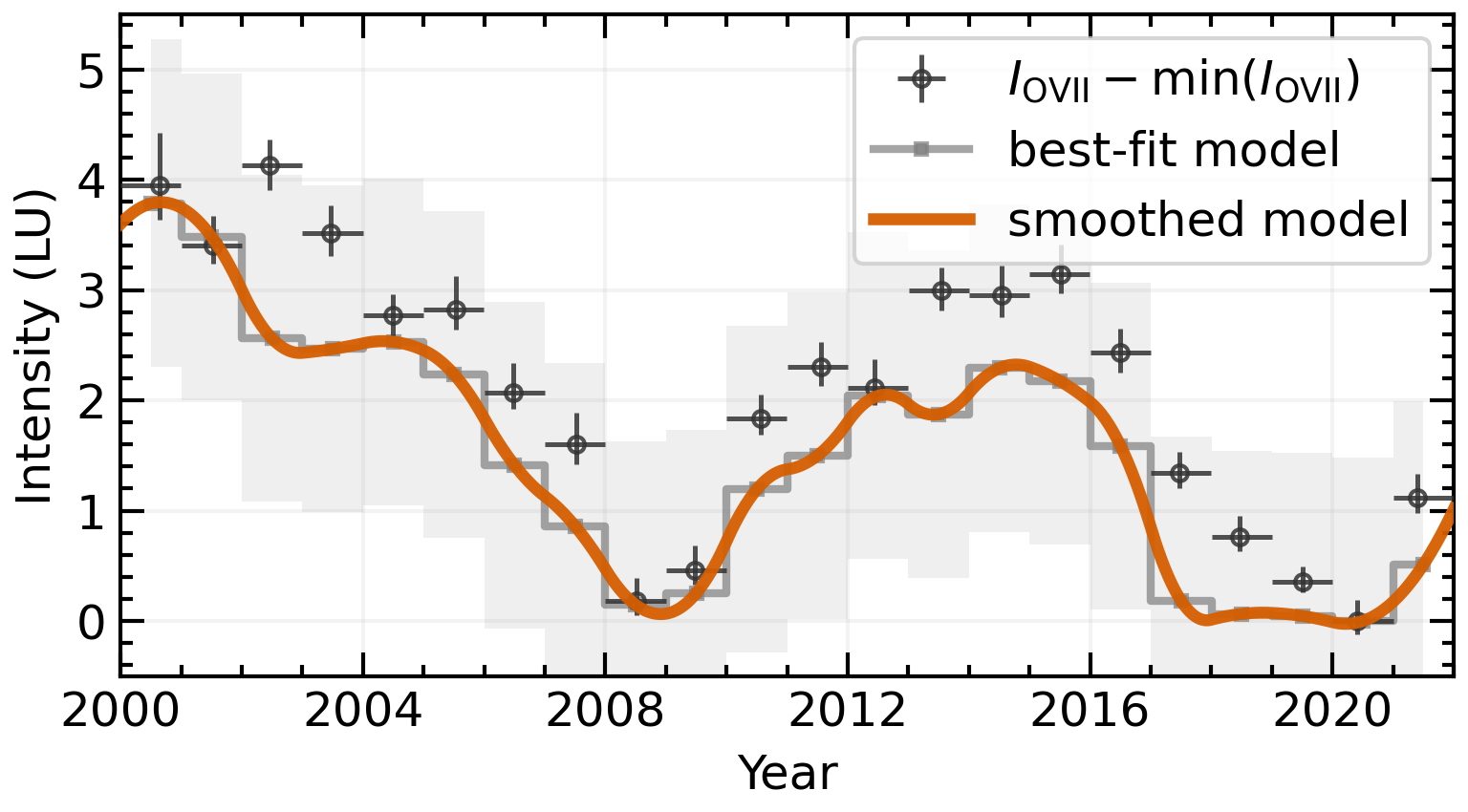}
	\hspace{0.02\textwidth}
	\includegraphics[width=0.465\textwidth]{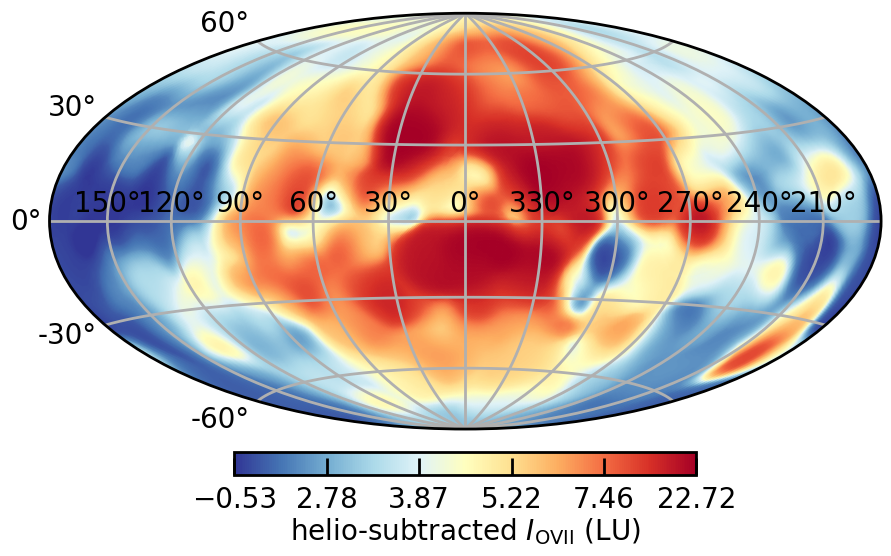}
	\caption{Removing heliospheric and MW hot gas contributions while retaining magnetospheric signals by using nightside observations ($X_{\rm GSE}<0$). Left: Long-term variation in the observed $I_{\rm OVII}$, attributed to heliospheric SWCX. This variation is characterized by a nonparametric model (smoothed line) using the “close-pair” method described in Section~\ref{app:helio}. Right: All-sky map of the heliospheric-SWCX-subtracted $I_{\rm OVII}$, primarily tracing the MW hot gas emission. The map is centered on the Galactic center and smoothed with a Gaussian kernel of $\sigma=5^\circ$.}
    \label{fig:fig2}
\end{figure*}

\subsection{Background Removal}
The observed $I_{\rm OVII}$ is comprised of multiple emission components, including the magnetospheric SWCX of interest, as well as heliospheric SWCX and MW hot gas emission. These latter two components act as strong contaminants that must be carefully removed to isolate the magnetospheric signal. Below, we describe their distinct properties and the methods used to subtract them from the data.

XMM-Newton's LOS is nearly perpendicular to the Sun--Earth line, so nightside sightlines ($X_{\rm GSE}<0$) predominantly probe space away from the dayside magnetosheath where magnetospheric SWCX is concentrated. Consequently, the magnetospheric contribution on the nightside is negligible. This allows us to use nightside observations as a clean baseline to characterize and remove the heliospheric SWCX and MW hot gas contributions.

\subsubsection{Heliospheric SWCX removal}
\label{app:helio}

Heliospheric SWCX exhibits long-term variations in the soft X-ray background and is strongly correlated with solar cycles \citep{Qu2022solar}. 
For the \ion{O}{7} line, this component has a median of $\sim$2~LU, accounting for $\sim$30\% of the total observed background on average \citepalias{Pan2024ApJS}.
Nevertheless, this component also exhibits spatial variations due to the non-uniform distribution of neutral interplanetary hydrogen and helium \citep{Lallement1984A&A...140..243L, Dalaudier1984A&A...134..171D}. Simulations show that this spatial variation contributes an average of $\sim$0.3~LU to $I_{\rm OVII}$ for \xmm observations \citep{Qu2022solar}. As its amplitude is relatively small compared to the temporal variation, this spatial dependence is not considered in the removal.

To characterize and remove the temporal variation, we apply the “close-pair” method described in \citetalias{Pan2024ApJS}. This approach compares observations taken in nearly the same direction but at different times. Because the MW hot gas emission is spatially correlated on small angular scales of $\lesssim 5^\circ$ \citep[e.g.,][\citetalias{Qu2024ApJ}]{Kaaret2020disk}, the intensity difference within each pair primarily reflects temporal variations in the heliospheric SWCX. 
Specifically, we identify observation pairs with angular separations less than $4^\circ$ and compute their intensity differences $\delta I_{\rm OVII}$. They are modeled as the differences of a step function with 22 free parameters, representing the heliospheric SWCX contribution in each year from 2000 to 2022. These parameters are fitted using the same MCMC approach introduced in \citetalias{Pan2024ApJS}.

The best-fit model and the $I_{\rm OVII}$ medians are shown in Figure \ref{fig:fig2}. The results show that the model largely captures the temporal variation of heliospheric SWCX. Because solar activity evolves gradually over time, we further smooth the step-wise model using quadratic interpolation to produce a continuous correction curve. This smoothed function provides a consistent estimate of the heliospheric SWCX intensity at any given time. 
We subtract the interpolated value corresponding to each observation date to remove the heliospheric component from the observed $I_{\rm OVII}$. 

The temporal model remains systematically below the annual medians, likely due to the uneven MW hot gas contribution introduced by non-uniform annual sky coverage.
To confirm this, we evaluate the impact of yearly variations in sky coverage on the MW \ion{O}{7} contribution. Since MW emission increases toward the Galactic center, we use the angular distance of each \xmm sightline to the Galactic center, $\theta_{\rm GC}$, as a proxy for the MW contribution \citep[e.g.,][\citetalias{Qu2024ApJ}]{Miller2015}. The result shows that the annual median $\theta_{\rm GC}$ varies between $\sim90^\circ$--$120^\circ$. In this range, the MW \ion{O}{7} intensity increases by $\sim$0.5 LU per 10$^{\circ}$ decrease in $\theta_{\rm GC}$ \citepalias{Qu2024ApJ}.
Moreover, the annual residual between the observed intensity and the temporal model is anti-correlated with $\theta_{\rm GC}$ (Pearson coefficient $r\approx-0.7$), confirming that the discrepancy originates primarily from the sampling-induced MW contribution. This contribution is subsequently removed using the empirical MW map in the following section.

\subsubsection{Galactic emission removal}
\label{app:MW}
The soft X-ray emission from MW hot gas is spatially extended and direction-dependent, with enhanced emissions toward the Galactic center. Large-scale features such as Superbubbles and supernova remnants further complicate its spatial distribution.

Utilizing this characteristic, we isolate the MW emission by constructing an empirical all-sky map using both the clean sample and the observations within the large-scale structures mentioned in Section~\ref{subsec:observation data}. Specifically, this map is generated by Gaussian-smoothing the heliospheric-SWCX–subtracted $I_{\rm OVII}$ with a kernel of $\sigma=5^\circ$ (Figure~\ref{fig:fig2}).
The MW contribution to each observation is then estimated from the map based on the pointing direction and subtracted accordingly. 

We cross-check our empirical MW map against the the field-by-field \ion{O}{7}intensities derived from HaloSat background spectra, obtained around solar minimum \citep{Bluem2022ApJ...936...72B}. To match these measurements, we extract the corresponding MW intensity from our map by calculating the median value within the FOV (radius $=5^\circ$) of each HaloSat field. The standard deviation of the intensity differences is $\sim$1.0 LU, comparable to the median measurement uncertainty (0.7 LU) of the \xmm $I_{\rm OVII}$ data, demonstrating that the two datasets are consistent within the measurement uncertainties.

\subsection{Magnetospheric SWCX Simulations} \label{subsec:simulation}
For each \xmm observation, we also extract the simulation-predicted magnetospheric SWCX signals, which will be compared directly with the observations in the following analysis.
In particular, the adopted magnetohydrodynamic (MHD) simulation follows \citet{Sun2021JGRA} and \citet{Yuqi2025FrASS..1263653G}, where the LOS-integrated SWCX collision rate, normalized by solid angle 
and \ion{O}{7} photon energy, is defined as:
\begin{equation}
\label{eq:SWCX_Q}
    Q^{\rm sim}_{\rm OVII} \equiv \frac{1}{4\pi E_{\rm OVII}} 
    \int n_{\rm H} n_{\rm SW} v_{\rm col} \ {\rm d}s,
\end{equation}
such that the predicted \ion{O}{7} line intensity (in units of LU) is simply:
\begin{equation}
\label{eq:SWCX_I}
    I^{\rm sim}_{\rm OVII} = \alpha^{\rm sim}_{\rm OVII} Q^{\rm sim}_{\rm OVII}
\end{equation}
where $E_{\rm OVII} \approx 0.56\ \rm keV$ is the \ion{O}{7} triplet line energy. The scaling factor $\alpha^{\rm sim}_{\rm OVII}$ represents the charge exchange efficiency for the \ion{O}{7} triplet, encompassing the relevant SWCX cross sections and the composition and abundances of the highly charged oxygen ions in the SW. 
Since the true value of $\alpha_{\rm OVII}$ is uncertain, we temporarily assume $\alpha^{\rm sim}_{\rm OVII} = 1 \times 10^{-15}$~eV~cm$^{2}$, a total soft X-ray efficiency commonly used in magnetospheric SWCX simulations (e.g., \citealt{Sun2019JGRA}), so that the simulated intensities are of the same order as the observations. Although the true $\alpha_{\rm OVII}$ is only a fraction of this total, this choice merely rescales the absolute intensity and does not affect the comparison of spatial morphology.
The empirical $\alpha_{\rm OVII}$ is later derived utilizing the simulated $Q^{\rm sim}_{\rm OVII}$ in Section~\ref{sec:Discussion and Summary}.

The exospheric hydrogen density $n_{\mathrm{H}}$ in $Q^{\rm sim}$ is modeled as $n_{\mathrm{H}} = 25\,\mathrm{cm}^{-3} \times (10\ \mathrm{R_E}/r)^3$ adopted from \citep{Cravens1997}. 
The SW proton density $n_{\rm SW}$ and the collision speed 
$v_{\rm col}$ (with $v_{\rm col}=\sqrt{v^2_{\rm SW}+v^2_{\rm th}}$ estimated from the SW bulk and thermal speed) are obtained from a global MHD model.

Although SW conditions vary in time, our MHD simulations show that the typical SW density variation (25th to 75th percentiles; assuming a fixed mean SW speed) results in about a $\sim\pm6\%$ shift in the magnetopause standoff distance. Given this structural stability, we adopt mean SW conditions ($n_{\rm SW}=5~{\rm cm^{-3}}$, $v_{\rm SW}=400~{\rm km~s^{-1}}$, $P=0.0126$ nPa, and IMF $B_x=B_y=0$ and $B_z=-5~{\rm nT}$) as model inputs, a configuration commonly used in magnetospheric MHD simulations \citep[e.g.,][]{Hu2007JGRA, Sun2015JGRA..120..266S}.
The resulting 3D plasma distribution provides the local SW density and velocity required for Equation~\ref{eq:SWCX_Q}.
The coupled MHD equations are solved numerically using the 3D extended Lagrangian version of the Piecewise Parabolic Method (PPMLR) MHD code developed by \citet{Hu2007JGRA}.
Using the model-predicted density and velocity outputs, we compute the emissivity and integrate it along each LOS following Equation \ref{eq:SWCX_Q}. The integration is confined to the region from outside the simulated magnetopause boundary out to $r = 80\ \rm R_E$, assuming negligible emission contributions beyond this distance (\citealt{Sun2019JGRA}).

\begin{figure*}[t!]
	\centering
	\includegraphics[width=0.475\textwidth]{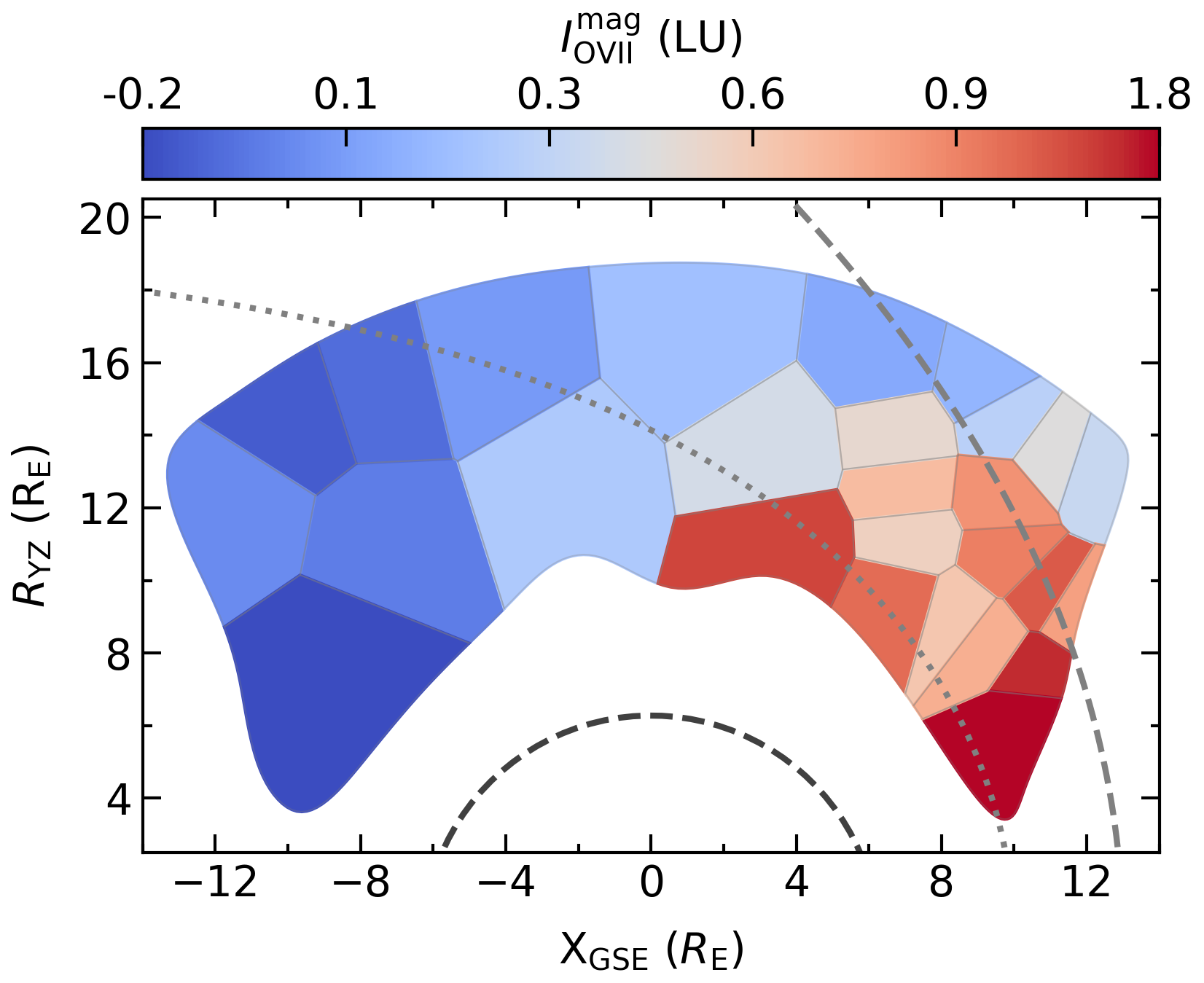}
	\hspace{0\textwidth}
	\includegraphics[width=0.475\textwidth]{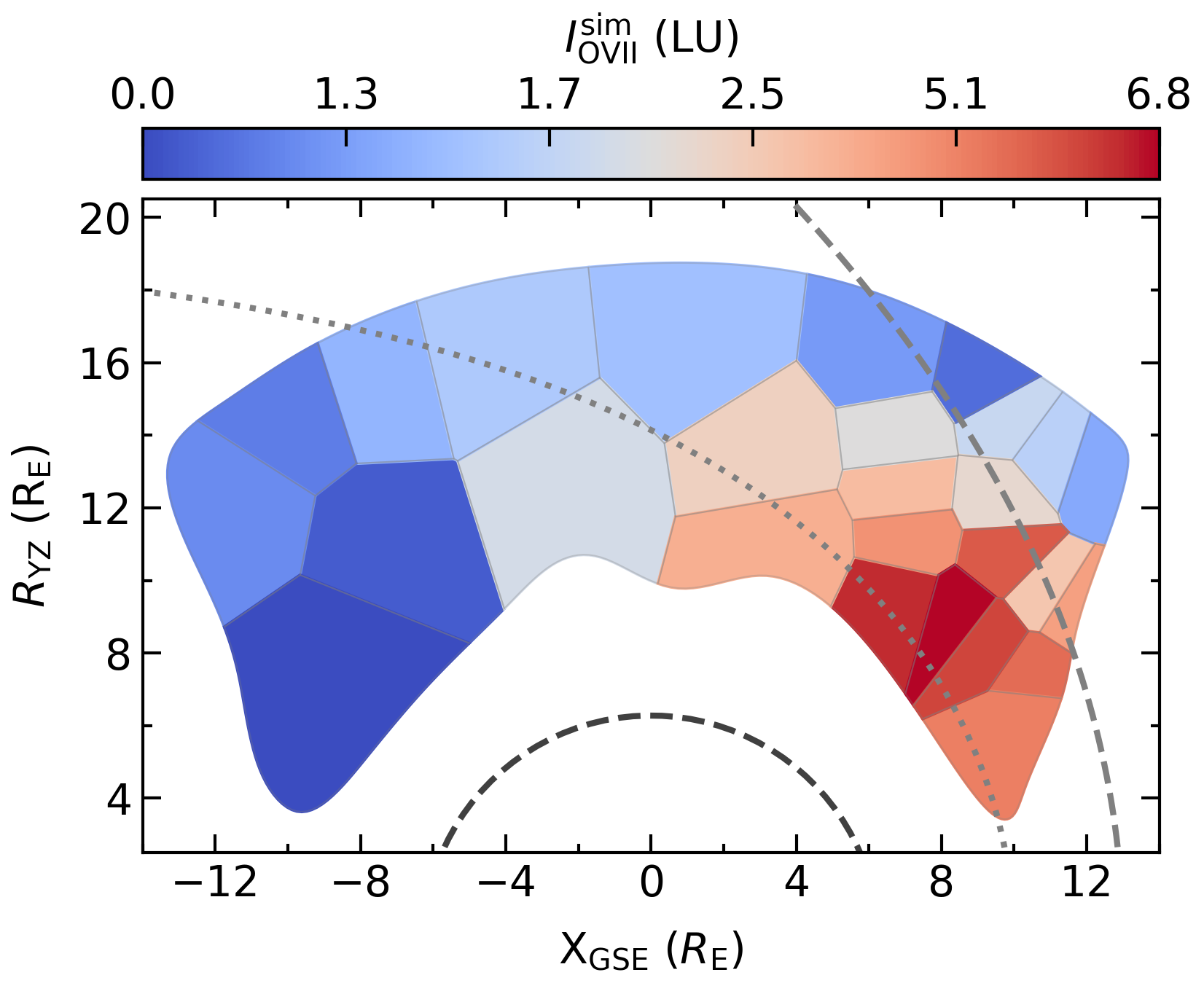}
    \caption{Spatial distribution of corrected \ion{O}{7} intensity ($I^{\rm mag}_{\rm OVII}$), compared with the predicted magnetospheric SWCX intensity ($I^{\rm sim}_{\rm OVII}$) from an MHD simulation (Section \ref{subsec:simulation}). The $I^{\rm mag}_{\rm OVII}$-enhanced region lies primarily within the modeled magnetosheath (between the dotted and dashed lines) and closely matches the simulated morphology, together confirming a magnetospheric SWCX origin. Both maps are adaptively binned to a fixed signal-to-noise ratio using the same Voronoi tessellation, with color representing the median intensity per bin. The $I^{\rm mag}_{\rm OVII}$ and $I^{\rm sim}_{\rm OVII}$ values in each bin maintain similar ratios, indicating a constant scaling factor and enabling the derivation of an empirical $\alpha_{\rm OVII}$.}
\label{fig:spatial}
\end{figure*}

\section{X-ray emission in magnetosphere}  \label{sec:magnetosphere}
To first order, Earth's magnetospheric structure is approximately axisymmetric along the Sun-Earth line (e.g., \citealt{Robertson2003GL016740}; for recent simulated emissivity maps, see \citealt{Yang2024E&PP....8...59Y, Xu2024JGRA..12932687X}).
This symmetry arises because the magnetosphere is primarily shaped by the highly supersonic SW flowing predominantly in the anti-sunward direction, which uniformly compresses Earth's dipolar magnetic field. 
Secondary asymmetries do exist, including a pole-plane (north--south) asymmetry caused by the geomagnetic dipole tilt and a dawn--dusk asymmetry resulting from SW aberration \citep[e.g.,][]{Zoennchen2013A}. However, neither is expected to significantly bias our results. The former is mitigated as the Earth avoidance constraint excludes most sightlines through the highly distorted polar cusps (see Section \ref{subsec:observation data}), while the latter is effectively averaged out across our broad spatial sample.

This large-scale structure includes three key components: (1) the bow shock, where the supersonic SW is first decelerated; (2) the magnetosheath, a region of shocked and compressed SW plasma; and (3) the magnetopause, the boundary where SW dynamic pressure balances the planetary magnetic pressure. 


Among these, the dayside magnetosheath is expected to exhibit the strongest SWCX emission due to enhanced ion–neutral interactions. \xmm observations confirm this, showing enhanced 0.5–0.7 keV flux when sightlines cross the dayside magnetosheath \citep[e.g.,][]{snowden2004ApJ...610.1182S, kuntz2008}. Statistical analysis further indicates that these enhanced observations are mostly distributed at $X_{\rm GSE}\approx 10\ \rm R_E$ \citep{Carter2011A&A}. In this section, we investigate the spatial distribution of magnetospheric SWCX and constrain its 3D structure and emissivity using the $I^{\rm mag}_{\rm OVII}$ data derived from 22 years of \xmm observations, covering two full, though unusual, solar cycles.

\begin{figure*}[t!]
\centering
\hspace*{-1cm}\includegraphics[width=0.92\textwidth]{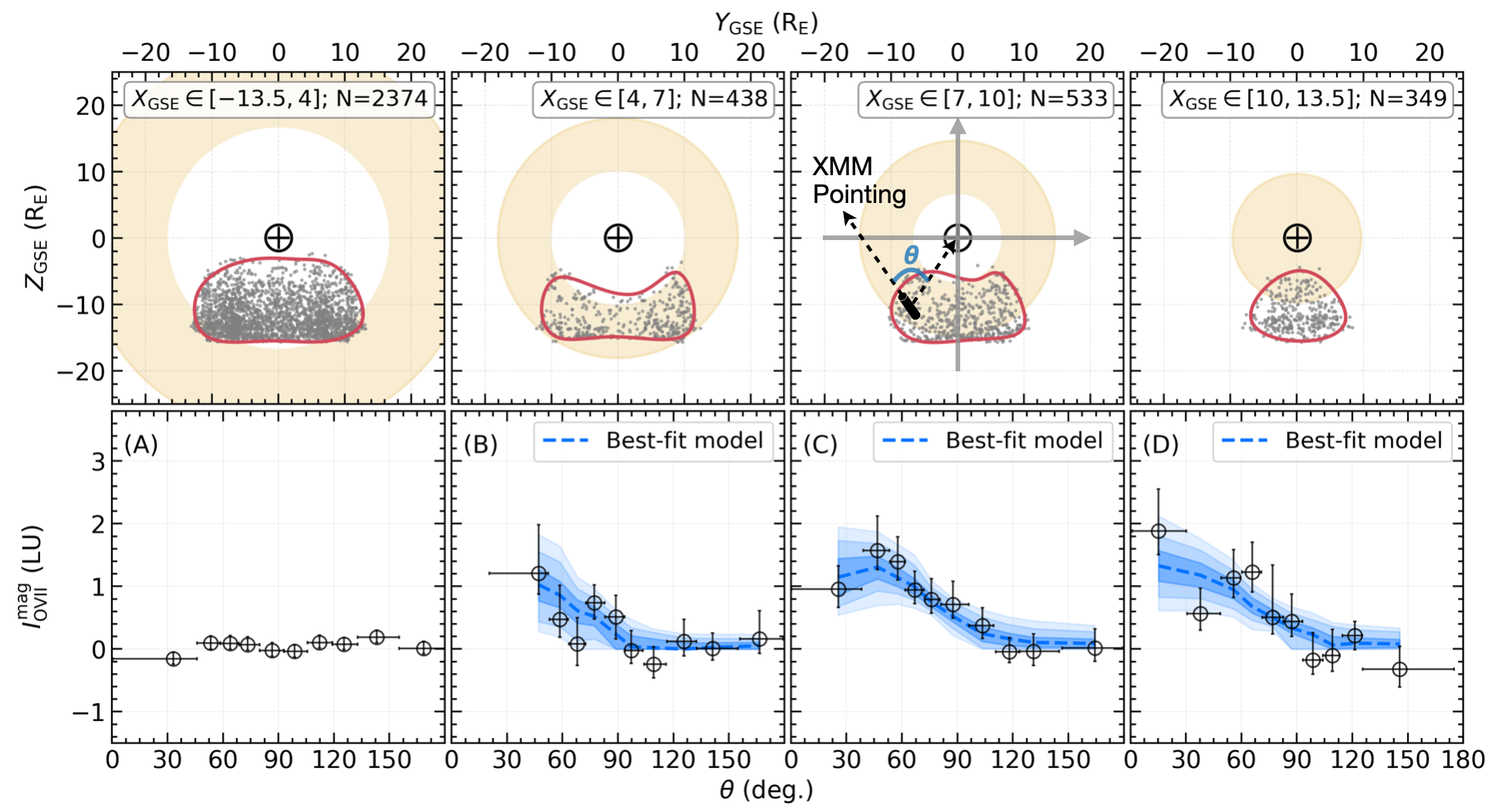}
\caption{Directional dependence of $I^{\rm mag}_{\rm OVII}$ demonstrating that longer magnetosheath path lengths correspond to stronger signals.
Top: Distribution of \xmm positions (enclosed by solid lines) in the $Y_{\rm GSE}$--$Z_{\rm GSE}$ plane across $X_{\rm GSE}$ slices. These positions sample the predicted magnetosheath regions (shaded shells), enabling a statistical study of directional effects. The third panel defines angle $\theta$, which quantifies the magnetosheath path length along the LOS; $\theta$ is the angle between the projection of the satellite's pointing direction onto the $Y_{\rm GSE}$--$Z_{\rm GSE}$ plane and the direction perpendicular to the $X_{\rm GSE}$ axis.
Bottom: $I^{\rm mag}_{\rm OVII}(\theta)$ profiles showing systematic variations with viewing direction. These profiles encode information about local magnetosheath structure and emissivity. An empirical model developed in this work successfully reproduces these profiles (shaded regions) and retrieves local structure and emissivity information (Section~\ref{sub:modeling}). Points represent bin medians with $1\sigma$ uncertainties (vertical error bars); horizontal bars indicate bin widths in $\theta$.}
\label{fig:Angle}
\end{figure*}

\subsection{Spatial Distribution}
\label{subsec:spatial} 
The \xmm orbits span dayside and nightside, covering the magnetosheath as shown in Figure \ref{fig1:data distribution}. 
This orbit configuration enables a direct probe of the spatial variation, as observations made within the magnetosheath are expected to capture higher SWCX signals.

Given the expected axisymmetry of the magnetospheric structure, the SWCX emission primarily varies with $X_{\rm GSE}$ and the radial distance in the $Y_{\rm GSE}$–$Z_{\rm GSE}$ plane ($R_{\rm YZ} \equiv \sqrt{Y_{\rm GSE}^2 + Z_{\rm GSE}^2}$).
To explore such spatial variation, we plot the $I^{\rm mag}_{\rm OVII}$ data at the median satellite positions in the $X_{\rm GSE}$--$R_{\rm YZ}$ plane. 
Note that for exposures exceeding $50~\rm ks$, XMM-Newton's highly elliptical orbit can cause the satellite position to vary by more than 3 and 4 $\rm R_E$ ($1\sigma$) in $X_{\rm GSE}$ and the radial direction, respectively. To ensure that the median position remains a reliable representation, we include only the data with exposures $<50~\rm ks$ in this analysis.
Given the weak signals and uneven sampling, we apply Voronoi tessellation \citep{Cappellari2003} with a target signal-to-noise ratio S/N = 11 to adaptively bin the data. This threshold typically results in bins larger than 1 $\rm R_E$ radius, mitigating the positional uncertainty introduced by the median-position approximation. 

The resulting spatial distribution (Figure~\ref{fig:spatial}) reveals a clear enhancement when the satellite is on the dayside, peaking at $X_{\rm GSE} \approx 10~\rm R_E$, where the intensity exceeds the minimum by $\approx$2.0 LU. The enhanced emission lies predominantly within the magnetosheath, defined by empirical magnetopause \citep{Shue1997} and bow shock \citep{Chao2002COSPA..12..127C} models for average SW conditions. 
We note that the actual S/N in each Voronoi bin (computed as $\text{S/N}_{\rm bin} = \sum S_i / \sqrt{\sum \sigma_i^2}$) is mostly below the target S/N = 11. This is because the input $I^{\rm mag}_{\rm OVII}$ data contain negative values after background subtraction, leading to partial signal cancellation within each bin. Here, we empirically quantify the significance of the dayside enhancement using the nightside residuals as a noise baseline. With a standard deviation of $\approx$1.7~LU and $>$30 data points per bin, the standard error of the mean is $\sim$0.3~LU, yielding a detection significance of $\sim$2--6 $\sigma$ in the enhanced dayside bins.

For comparison, the simulation-predicted magnetospheric SWCX ($I_{\rm OVII}^{\rm sim}$) is rebinned using the same Voronoi tessellation and shown in Figure~\ref{fig:spatial}.
The predicted magnetospheric SWCX largely overlaps with the observation. 
Its peak intensity occurs within $X_{\rm GSE}=8$--$10~\mathrm{R_E}$, consistent with the observed peak near $10~\mathrm{R_E}$. These agreements collectively support a magnetospheric SWCX origin for the emission.

Note that Figure~\ref{fig:spatial} is not equivalent to a true 3D emissivity map, but rather represents the integrated intensity at the median satellite positions. This representation mixes sightlines that intersect the magnetosheath with those that do not, and artificially associates the emission with the spacecraft location rather than its physical source region. These limitations are addressed by directional and 3D modeling in the following subsections.

\subsection{Directional Dependence}
\label{subsec:angle}
The 2D intensity distribution (Figure \ref{fig:spatial}) inherently encodes the underlying 3D emissivity structure (i.e., the geometric structure and emissivity distribution). 
Inferring the 3D structure requires the satellite's position and direction, as the measured intensity depends on the path length through the magnetosheath and the local emissivity along that path.
Here, we investigate the 3D magnetosheath structure assuming an axisymmetry around the $X_{\rm GSE}$ axis.
This symmetry implies that at any given $X_{\rm GSE}$, the magnetosheath forms a shell-like cross-section in the $Y_{\rm GSE}$–$Z_{\rm GSE}$ plane bounded between the local magnetopause and bow shock radii ($R_{\rm MP}$ and $R_{\rm BS}$). 
Then, the global structure is determined by these two radii as functions of $X_{\rm GSE}$.

Under this geometry, the \xmm pointing direction determines the path length between $R_{\rm MP}$ and $R_{\rm BS}$ in the local $Y_{\rm GSE}$–$Z_{\rm GSE}$ plane at a given $X_{\rm GSE}$.
Therefore, a directional dependence of the magnetospheric SWCX is expected, and 
can be parameterized by the angle $\theta$, which is defined in that plane between the \xmm LOS and the direction from the satellite to the plane's center (see the illustration in Figure~\ref{fig:Angle}). 
The observed intensity $I(\theta)$ can be modeled as
\begin{equation}
I(\theta) = \epsilon \cdot s(\theta| R_{\rm MP}, R_{\rm BS}),
\label{eq:model}
\end{equation}
where $\epsilon$ is the local emissivity and $s$ is the path length through the shell at a given $X_{\rm GSE}$. 
The SW density within the magnetosheath is non-uniform, typically peaking near the magnetopause and declining toward the bow shock, leading to spatial variations in emissivity. As a zero-order approximation, we assume a constant emissivity within each local shell between $R_{\rm MP}$ and $R_{\rm BS}$.

Under this simplification, for a fixed satellite position, the variation in $I(\theta)$ is solely determined by the path length through the magnetosheath.
As $\theta$ increases from zero, the path length first rises as the LOS cuts through more of the shell.
This rise may be absent near the subsolar point, where the magnetopause radius is small. 
As $\theta$ further increases, the path length peaks when the LOS is tangent to the magnetopause, followed by a decrease to zero when the LOS becomes tangent to the bow shock. 
Consequently, this geometry predicts a distinct rise–peak–fall pattern in $I(\theta)$ for magnetosheath SWCX emission. 
Detecting this pattern in our $I^{\rm mag}_{\rm OVII}(\theta)$ data would both confirm the directional signature of the emission and further constrain the 3D structure of the magnetosheath.

In the following, we examine the directional dependence in the $I_{\rm OVII}^{\rm mag}$.
Because all \xmm observations are pointed nearly perpendicular to the $X_{\rm GSE}$ axis within $\pm 20^\circ$ (Section~\ref{subsec:observation data}), the projected path length in the $Y_{\rm GSE}$--$Z_{\rm GSE}$ plane differs from the real path length along any LOS by $\lesssim 6\%$ (i.e., $1-\cos 20^\circ$).
This small deviation justifies using $\theta$ defined in the $Y_{\rm GSE}$--$Z_{\rm GSE}$ plane for our analysis.

To ensure sufficient S/N for the directional dependence function, we divide our data into four $X_{\rm GSE}$ slices (panels A–D in Figure \ref{fig:Angle}), spanning from the nightside to the dayside. Slice A ($[-13.5,4]~\rm R_E$) represents a weak emission region, serving as a baseline. 
The SWCX-enhanced region ($[4, 13.5]~\rm R_E$) is further divided into slices B ($[4, 7]~\rm R_E$), C ($[7, 10]~\rm R_E$), and D ($[10, 13.5]~\rm R_E$). 
We construct the $I^{\rm mag}_{\rm OVII}(\theta)$ profile within each $X_{\rm GSE}$ slice by sorting the $\theta$ values into 10 equal-number bins.
In each bin, the median intensity and $1\sigma$ uncertainties are computed from the merged MCMC posterior samples of all contributing observations within that bin.

The resulting $I^{\rm mag}_{\rm OVII}(\theta)$ profiles (Figure~\ref{fig:Angle}) confirm the predicted directional dependence. The dayside profiles generally follow the expected rise-peak-fall pattern, most evident in Slice C. 
Slice B lacks a clear rise, likely due to limited coverage at small $\theta$, while Slice D shows no initial rise because it samples the subsolar magnetosheath, where the maximum path length occurs at $\theta=0^{\circ}$. 
Consequently, their profiles peak at the smallest $\theta$ available and then fall as $\theta$ increases.
By contrast, Slice A shows no obvious directional dependence, primarily because these sightlines scan the flanks, where the emissivity is negligible. Furthermore, Slice A observations are located inside the magnetopause, causing the LOS path lengths to vary only minimally with $\theta$, further suppressing any detectable angular variation in $I^{\rm mag}_{\rm OVII}(\theta)$. 
These results confirm the magnetosheath's emissivity structure and reveal that the $I^{\rm mag}_{\rm OVII}(\theta)$ profiles encode information about its geometry and emissivity.

\begin{figure*}[t!]
    \centering
    \begin{minipage}[t]{0.44\textwidth}
        \centering
        \includegraphics[width=\linewidth]{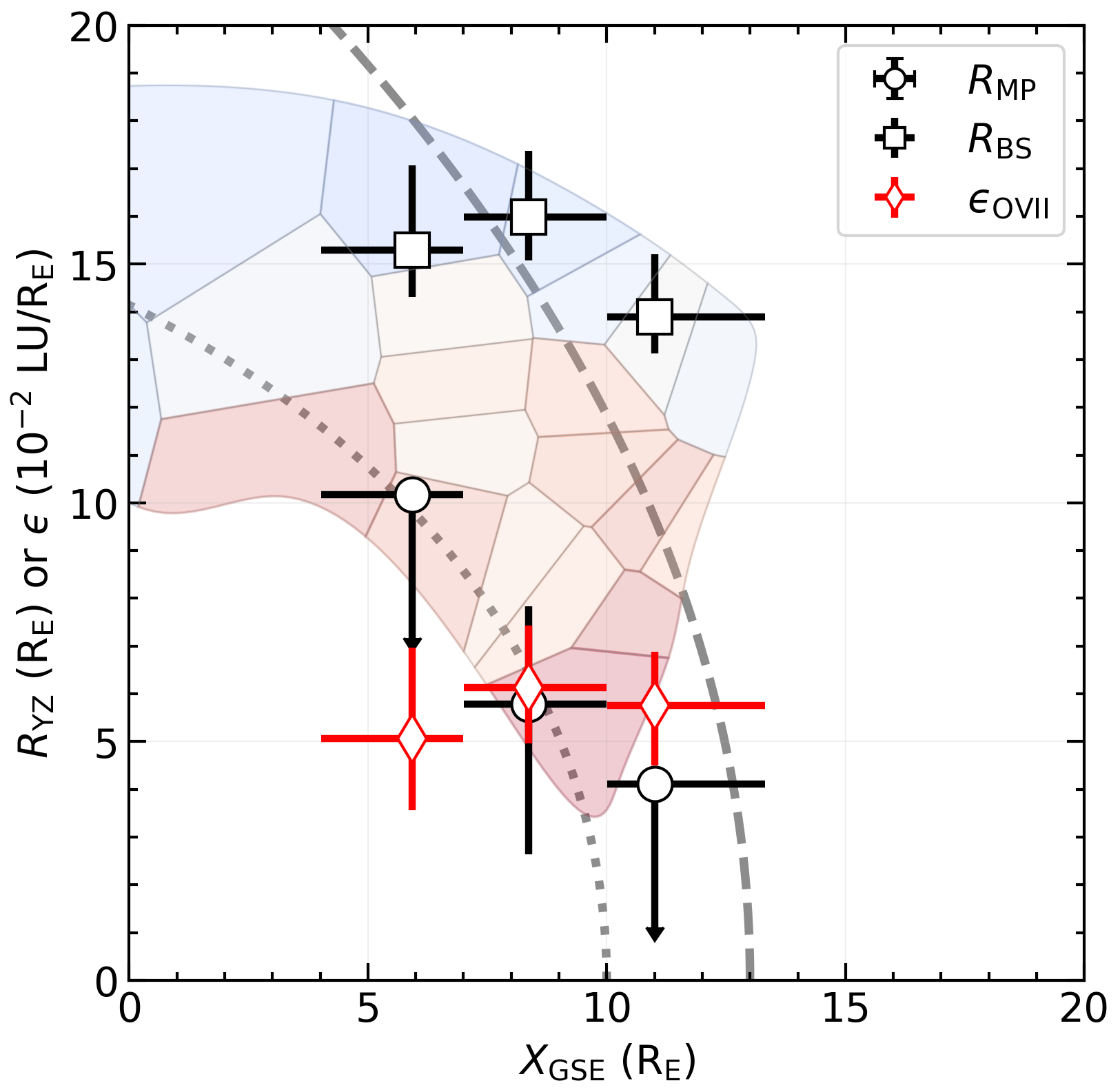}
    \end{minipage}
    \hspace{0.02\textwidth}
    \begin{minipage}[t]{0.44\textwidth}
        \centering
        \includegraphics[width=\linewidth]{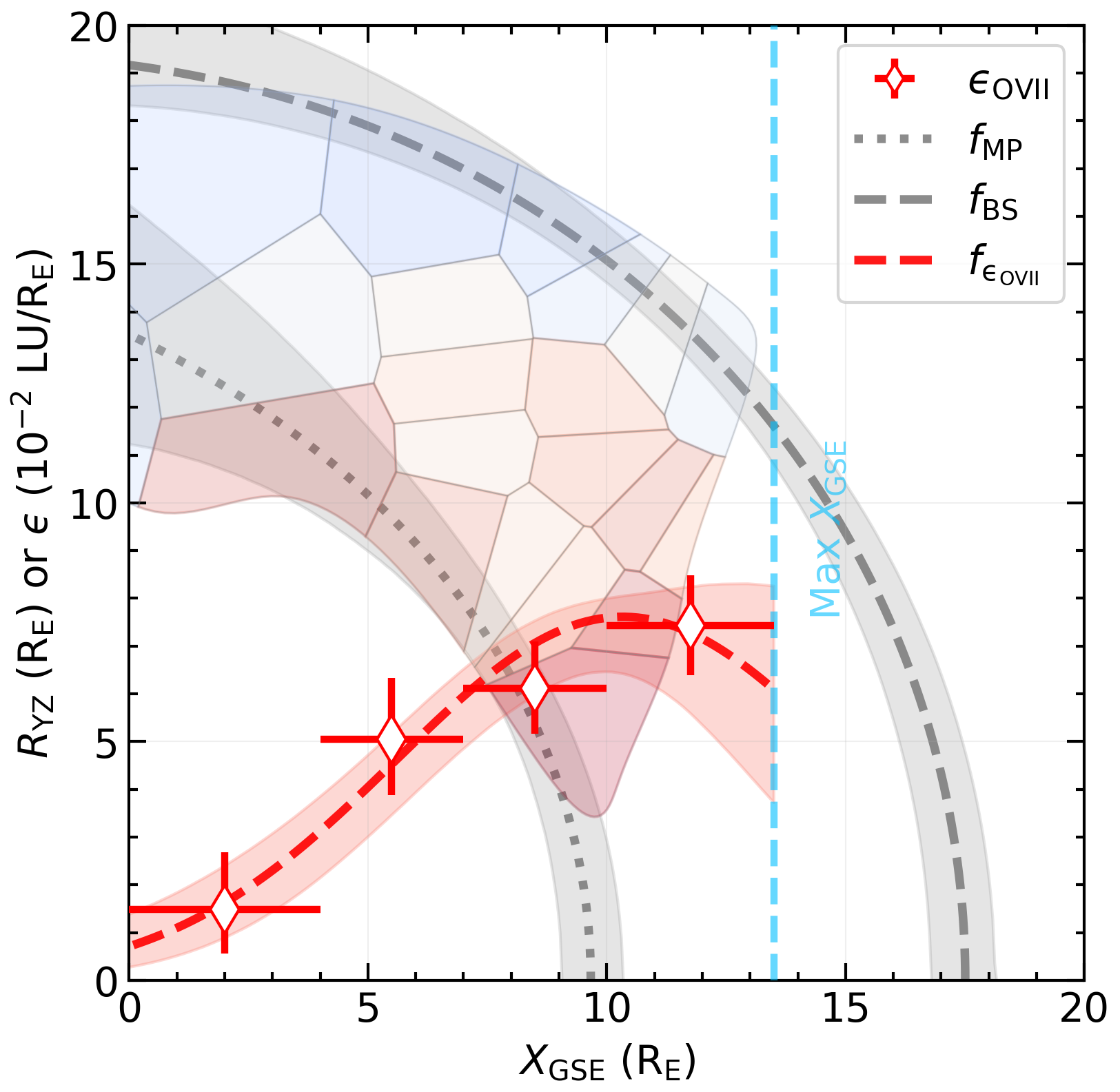}
    \end{minipage}
\caption{Magnetopause and bow shock boundaries ($R_{\rm MP}$ and $R_{\rm BS}$) derived from $X_{\rm GSE}$ slices, overlaid on the spatial distribution from Figure~\ref{fig:spatial}. These boundaries closely trace the region of strong emission, demonstrating that they reasonably capture the expected magnetosheath structure.
Right: Best-fit functions describing the magnetosheath structure and the SWCX emissivity within it. The $R_{\rm MP}$ and $R_{\rm BS}$ boundaries are parameterized using the functional form proposed by \citet{Shue1997}, which the boundaries appear to follow. The emissivity profile is modeled by a Gaussian function.}
\label{fig5:2d3dshape}
\end{figure*}

\subsection{An Empirical Model of the Magnetosphere}
\label{sub:modeling}

Building on these directional dependences, we derive an empirical model of the magnetosheath averaged over 22 years.
Here, we adopt a two-step approach: (1) simplified 2D models to verify whether the data can resolve the magnetosheath structure, and (2) a more rigorous axisymmetric 3D model after validation.

First, we model the magnetosheath in each slice as a constant-emissivity shell described by Equation~\ref{eq:model}. The three parameters ($\epsilon$, $R_{\rm MP}$, $R_{\rm BS}$) are determined by fitting this model to the $I^{\rm mag}_{\rm OVII}(\theta)$ profile under a Bayesian framework. The likelihood is defined as
\begin{equation}
\ln p = -\frac{1}{2} \sum \frac{(I - I_{\rm m})^2}{\sigma_I^2},
\end{equation}
where $I$ and $\sigma_I$ are the median intensity and its uncertainty for each bin, and $I_{\rm m}$ is the model prediction. We adopt uniform priors requiring all parameters to be positive and perform the fitting using the \texttt{emcee} package \citep{foreman2013emcee}. The best-fit parameters are reported in Table~\ref{tab:boundary_fit}.

\begin{table}[t]
\centering
\caption{Best-fit parameters derived from modeling the $I^{\rm{mag}}_{\rm OVII}(\theta)$ profiles.}
\begin{threeparttable} 
\setlength{\tabcolsep}{0.2cm}
\begin{tabular*}{\linewidth}{@{\extracolsep{\fill}}cccc}  
\toprule
\multicolumn{1}{c}{$X_{\rm GSE}$ Range} & 
\multicolumn{1}{c}{$R_{\rm MP}$} & 
\multicolumn{1}{c}{$R_{\rm BS}$} & 
\multicolumn{1}{c}{$\epsilon$ (10$^{-2}$ LU/$\rm R_E$)} \\
\midrule
$[4.0, 7.0]$   & $<10.2$ & $15.3^{+1.8}_{-1.0}$ & $5.1^{+1.9}_{-1.5}$ \\
$[7.0, 10.0]$  & $5.8^{+2.0}_{-3.2}$ & $16.0^{+1.4}_{-0.9}$ & $6.1^{+1.3}_{-1.2}$ \\
$[10.0, 13.5]$ & $<4.1$ & $13.9^{+1.3}_{-0.8}$ & $5.8^{+1.1}_{-1.3}$ \\
\bottomrule
\end{tabular*}
\begin{tablenotes}[flushleft]
    \small
    \item[] \textit{Note.} For well-constrained parameters, we report the posterior median with asymmetric $1\sigma$ uncertainties from the MCMC posterior distribution. For unconstrained parameters, we report the 95\% credible upper limit.
\end{tablenotes}
\end{threeparttable}
\label{tab:boundary_fit}
\end{table}

The best-fit models successfully reproduce the 
characteristic rise--peak--fall pattern across Slices B--D (Figure~\ref{fig:Angle}, bottom panels), validating the 
constant-emissivity shell geometry within each $X_{\rm GSE}$ slice. The derived boundaries follow the \citet{Shue1997} functional form, with the emissivity peaking near the subsolar magnetosheath, consistent with enhanced SW--neutral interactions on the dayside (Figure~\ref{fig5:2d3dshape}).

\begin{table*}[t]
\centering
\caption{Best-fit parameters derived from modeling the 3D magnetosheath.}
\begin{threeparttable}
\setlength{\tabcolsep}{0.2cm}
\begin{tabular*}{\textwidth}{@{\extracolsep{\fill}}ccccc}
\toprule
\multicolumn{1}{c}{Geometry} &
\multicolumn{1}{c}{$r_0~(\rm R_E)$} &
\multicolumn{1}{c}{$a$} &
\multicolumn{1}{c}{$r_0^*~(\rm R_E)$} &
\multicolumn{1}{c}{$a^*$} \\
\midrule
Magnetopause & $9.7^{+0.7}_{-0.6}$  & $0.5^{+0.3}_{-0.3}$ & $10.3$ & $0.5$ \\
Bow shock    & $17.5^{+0.6}_{-0.7}$ & $<0.5$               & $13.0$ & $0.8$ \\
\midrule
\multicolumn{1}{c}{Emissivity} &
\multicolumn{1}{c}{$\epsilon~(10^{-2}~\rm LU/R_E)$} &
\multicolumn{1}{c}{$\alpha_\epsilon~(10^{-16}~\rm eV~cm^2)$} &
\multicolumn{1}{c}{$\epsilon^*~(10^{-2}~\rm LU/R_E)$} &
\multicolumn{1}{c}{$\alpha^*_\epsilon~(10^{-16}~\rm eV~cm^2)$} \\
\midrule
$\epsilon_1$ & $1.5^{+1.2}_{-0.9}$  & $1.5^{+1.2}_{-0.9}$ & $1.3^{+0.9}_{-0.8}$  & $1.3^{+0.9}_{-0.8}$ \\
$\epsilon_2$ & $5.0^{+1.3}_{-1.2}$  & $1.9^{+0.5}_{-0.5}$ & $5.2^{+1.1}_{-1.1}$  & $2.0^{+0.4}_{-0.4}$ \\
$\epsilon_3$ & $6.1^{+1.0}_{-0.9}$  & $1.6^{+0.3}_{-0.2}$ & $8.0^{+1.2}_{-1.2}$  & $2.1^{+0.3}_{-0.3}$ \\
$\epsilon_4$ & $7.4^{+1.0}_{-1.1}$  & $1.6^{+0.2}_{-0.2}$ & $13.3^{+2.3}_{-2.1}$ & $2.9^{+0.5}_{-0.5}$ \\

\bottomrule
\end{tabular*}
\begin{tablenotes}[flushleft]
    \small
    \item[] \textit{Note.} Starred quantities are obtained by fixing the MP and BS boundaries derived from our MHD simulation (Section~\ref{sub:modeling}), and fitting only the emissivities. The effective efficiency $\alpha_\epsilon$ is converted from the emissivity using $\alpha_\epsilon=\alpha^{\rm sim}_{\rm OVII}\,\epsilon/\epsilon_{\rm MHD}$, where $\epsilon_{\rm MHD}$ represents the MHD-predicted mean volume emissivity in each $X_{\rm GSE}$ slice.
\end{tablenotes}
\end{threeparttable}
\label{tab:3dfit}
\end{table*}

These results suggest that the XMM data are capable of resolving the magnetosheath structure, motivating a more rigorous 3D global model.
In this 3D framework, the magnetosheath is modeled as the volume enclosed between the MP and BS, each described by the \citet{Shue1997} parametrization:
\begin{equation}
r(\phi) = r_0 \left(\frac{2}{1+\cos\phi}\right)^a,
\label{Shuemodel}
\end{equation}
where the radial distance $r$ from Earth to the boundary is a function of the angle $\phi$ measured from the positive $X_{\rm GSE}$ axis ($\phi = 0^\circ$ at the subsolar point, $\phi = 180^\circ$ in the anti-sunward direction). Here, $r_0$ is the standoff distance at the subsolar point, and $a$ controls the flaring of the boundary.

Inside the magnetosheath, the SWCX emissivity is assumed to be uniform within discrete segments along $X_{\rm GSE}$:
\begin{equation}
\epsilon(X) = \epsilon_k, \quad 
X \in [X_k, X_{k+1}],
\end{equation}
with boundaries at $X_{\rm edges} = [0, 4, 7, 10, 13.5] \ \rm R_E$, consistent with the 2D slice intervals for a direct comparison between the two methods.

Each LOS is discretized into segments of $0.05~\rm R_E$, extending to $60\ \rm R_E$ from the satellite position to ensure complete magnetosheath coverage. 
With the boundary geometry and emissivity defined above, the modeled intensity is computed by summing contributions from segments traversing the magnetosheath:
\begin{equation}
I_{\rm model} = \int_{\rm LOS} \epsilon \, {\rm d}s 
\approx \sum_{k} \epsilon_k \Delta s_k,
\end{equation}
where $\Delta s_k$ is the 3D path length within the $k$-th $X_{\rm GSE}$ interval.

The model thus has eight free parameters $(r_0^{\rm MP}, a^{\rm MP}, r_0^{\rm BS}, a^{\rm BS}, \epsilon_1, \epsilon_2, \epsilon_3, \epsilon_4)$. We adopt uniform priors for all parameters, with physically motivated lower bounds: $r_0^{\rm MP} > 7\ \rm R_E$, corresponding to the most compressed magnetopause predicted by the \citet{Shue1997} relation under the extreme SW conditions, and $a^{\rm MP}, a^{\rm BS} > 0$ to ensure the boundaries flare outward within the dayside fitting region ($X_{\rm GSE} > 0$).
Before fitting, the $I^{\rm mag}_{\rm OVII}$ data are further filtered to exclude points with $I^{\rm mag}_{\rm OVII}>2\sigma$ of the sample mean. This filtering excludes 45 data points that are mainly located near Galactic bubble edges, incompletely excluded in Section~\ref{subsec:observation data}. 
These parameters are then determined by fitting the model to the remaining dayside $I^{\rm mag}_{\rm OVII}$ data under the Bayesian framework described in Section~\ref{sub:modeling}. 

To assess possible degeneracy between the boundary geometry and the emissivity normalization, we also perform an emissivity-only fit with the MP and BS boundaries fixed to those inferred from our MHD simulation. Specifically, the boundary locations are identified from radial emissivity gradients in the $X_{\rm GSE}$-$Y_{\rm GSE}$ plane ($Z_{\rm GSE}=0$) and fitted with the Shue profile (Equation~\ref{Shuemodel}), giving $r_0^{\rm MP}, a^{\rm MP}, r_0^{\rm BS}, 
a^{\rm BS} = 10.3~{\rm R_E},\ 0.5,\ 13.0~{\rm R_E},\ 0.8$. The best-fit parameters are reported in Table~\ref{tab:3dfit}.

The derived 3D magnetosheath boundary closely encloses the region of enhanced $I^{\rm mag}_{\rm OVII}$ emission (Figure~\ref{fig5:2d3dshape}), capturing the magnetospheric SWCX signal seen in the \xmm data. The results are consistent with the boundary radii derived from the 2D slice fitting, confirming the reliability of the 3D modeling approach.

The derived MP shape is highly consistent with theoretical and empirical expectations. The fitted standoff distance and flaring parameter closely match both our MHD simulation and the \citet{Shue1997} prediction under the mean SW conditions over the analyzed observation period ($n_{\rm SW} = 5.5~\rm cm^{-3}$, $v_{\rm SW} = 420~\rm km~s^{-1}$). Specifically, the fitted $r^{\rm MP}_{0} = 9.7^{+0.7}_{-0.6}~\rm R_{E}$ agrees within $1~\rm R_{E}$ with the expected value of $10.3~\rm R_{E}$ from both models, while the $a^{\rm MP}=0.5^{+0.3}_{-0.3}$ aligns well with the predicted values of $\sim$0.5.

In contrast, the BS parameters are poorly constrained due to a combination of physical and observational limitations. Unlike the MP, the BS lacks a sharp boundary because the X-ray emissivity decays gradually outward. This intrinsic diffuseness makes its overall shape difficult to determine from X-ray data alone, leading to large uncertainties in $a^{\rm BS}$. Furthermore, the dataset lacks spatial coverage near the subsolar region, which is essential for anchoring $r^{\rm BS}_{0}$. 
As a mere extrapolation rather than a direct measurement, it is unsurprising that the fitted $r^{\rm BS}_{0} = 17.5^{+0.6}_{-0.7}~\rm R_E$ shows a discrepancy with the expected values of 13.0 and 13.4 $\rm R_E$ from our MHD simulation and the \citet{Chao2002COSPA..12..127C} prediction, respectively.

The fitted emissivities $\epsilon_1$--$\epsilon_4$ are highly consistent with the 2D results, though the 3D model yields a slightly higher $\epsilon_4$. This is likely due to the binning operations in the 2D approach that tend to smooth out local intensity variations, thereby lowering the emissivity estimates.
By avoiding these binning effects, the 3D model provides a more direct description of the mean magnetospheric SWCX emission.

The treatment of boundary parameters, whether as free variables or fixed to MHD-derived values, significantly affects the derived emissivity (Table~\ref{tab:3dfit}). 
Adopting the MHD-derived boundaries increases the emissivities in the outer two dayside slices, with $\epsilon_4$ nearly doubling from $7.4^{+1.0}_{-1.1}$ to $13.3^{+2.3}_{-2.1}\times10^{-2}~{\rm LU/ R_E}$. This increase occurs because the MHD-inferred BS has a smaller standoff distance, leading to shorter modeled LOS path lengths through the dayside magnetosheath and therefore requiring a higher emissivity normalization to reproduce the observed intensity. This strong dependence highlights the need for future X-ray missions, such as SMILE \citep{SMILE2025SSRv..221....9W}, to better resolve the BS boundary, thereby placing stronger constraints on the emissivity.

Physically, the emissivity is expected to increase along $X_{\rm GSE}$, peak near the subsolar MP ($r^{\rm MP}_{0}$) where both exospheric hydrogen density and SW ion density are enhanced, and then decrease. Since $\epsilon_1$--$\epsilon_4$ only provide discrete averages within discrete $X_{\rm GSE}$ intervals, we fit these values with a Gaussian function to reconstruct this expected continuous profile:
\begin{equation}
\epsilon(X) = \epsilon_0 \exp\left(-\frac{(X - X_0)^2}{2\sigma_{ X}^2}\right),
\end{equation}
where $\epsilon_0$ represents the peak emissivity in subsolar magnetosheath at $X_{\rm GSE}=X_0$, and $\sigma_{X}$ controls the profile width. 

We employ a bootstrap approach to properly propagate the full uncertainty into the fitting. 
In each iteration, we draw random $X_{\rm GSE}$ and $\epsilon_{\rm bs}$ for each slice $[X_k, X_{k+1}]$, assembling a bootstrap dataset $(X_{\rm bs}, \epsilon_{\rm bs})$. In particular, $X_{\rm bs}$ is determined as the median of a bootstrapped $X_{\rm GSE}$ sample with replacements, while $\epsilon_{\rm bs}$ is directly extracted from the posterior distribution of the 3D model fit. We then perform a least-squares fit of the model parameters $(\epsilon_0, X_0, \sigma_X)$ by minimizing:
\begin{equation}
\chi^2 = \sum \left[\epsilon_{{\rm bs},i} - \epsilon_{\rm model}(X_{{\rm bs},i}|\epsilon_0, X_0, \sigma_X)\right]^2.
\end{equation}
The best-fit parameters are $\epsilon_0 = 7.6^{+1.2}_{-0.9} \times 10^{-2}~\rm LU/ R_E$, $X_0 = 10.3^{+2.9}_{-1.3}~\rm R_E$, $\sigma_{X} = 4.7^{+2.0}_{-1.0}\rm~ R_E$, and the modeled emissivity profile $\epsilon(X)$ is shown in Fig \ref{fig5:2d3dshape}.

This profile is compared with the 3D magnetosheath emissivity model of \citet{Jorgensen2019JGRA..124.4365J}. The model is an analytical function describing the total SWCX emission as a function of position in the magnetosheath, approximating the emissivity distribution derived from MHD simulations. The simulation is performed under dense SW conditions ($n_{\rm SW} = 22.5~\rm cm^{-3}$, $v_{\rm SW} = 400~\rm km~s^{-1}$), with $n_{\rm SW}$ about 4 times higher than the typical mean of $5.5~\rm cm^{-3}$, to produce a clearer emissivity structure. For a direct comparison, we average their model over the $Y_{\rm GSE}$–$Z_{\rm GSE}$ plane to obtain the corresponding profile $\epsilon_{\rm model}(X)$ and fit it with the same Gaussian function used to derive $\epsilon(X)$. Because the model predicts the total SWCX emission (not specifically the \ion{O}{7} line), we focus on comparing morphological trends rather than absolute amplitudes.

The $\epsilon_{\rm model}(X)$ shows a gradual increase, a peak, and a subsequent decline along $X_{\rm GSE}$, a trend similar to the derived profile. Specifically, the fitted peak center and width of $\epsilon_{\rm model}(X)$ are approximately $8.7~\mathrm{R_E}$ and $3.0~\mathrm{R_E}$, both smaller than the derived $X_0 = 10.3^{+2.9}_{-1.3}~\rm R_E$ and $\sigma_{X} = 4.7^{+2.0}_{-1.0}\rm~ R_E$. 
These discrepancies likely arise from the high SW density adopted in \citet{Jorgensen2019JGRA..124.4365J}. The enhanced dynamic pressure compresses the magnetosheath earthward, shifting the emissivity peak to smaller $X_{\rm GSE}$ and narrowing its extent along the Sun–Earth line. Nevertheless, the overall morphological agreement supports the reliability of our derived emissivity.

Together with the magnetosheath boundaries, the 3D model estimates the mean magnetospheric SWCX intensity under static SW conditions for any LOS passing through the magnetosheath within the data range ($X_{\rm GSE} \in [0, 13.5]~\rm R_E$).


\begin{figure*}[t!]
\centering
\includegraphics[width=0.8 \textwidth]{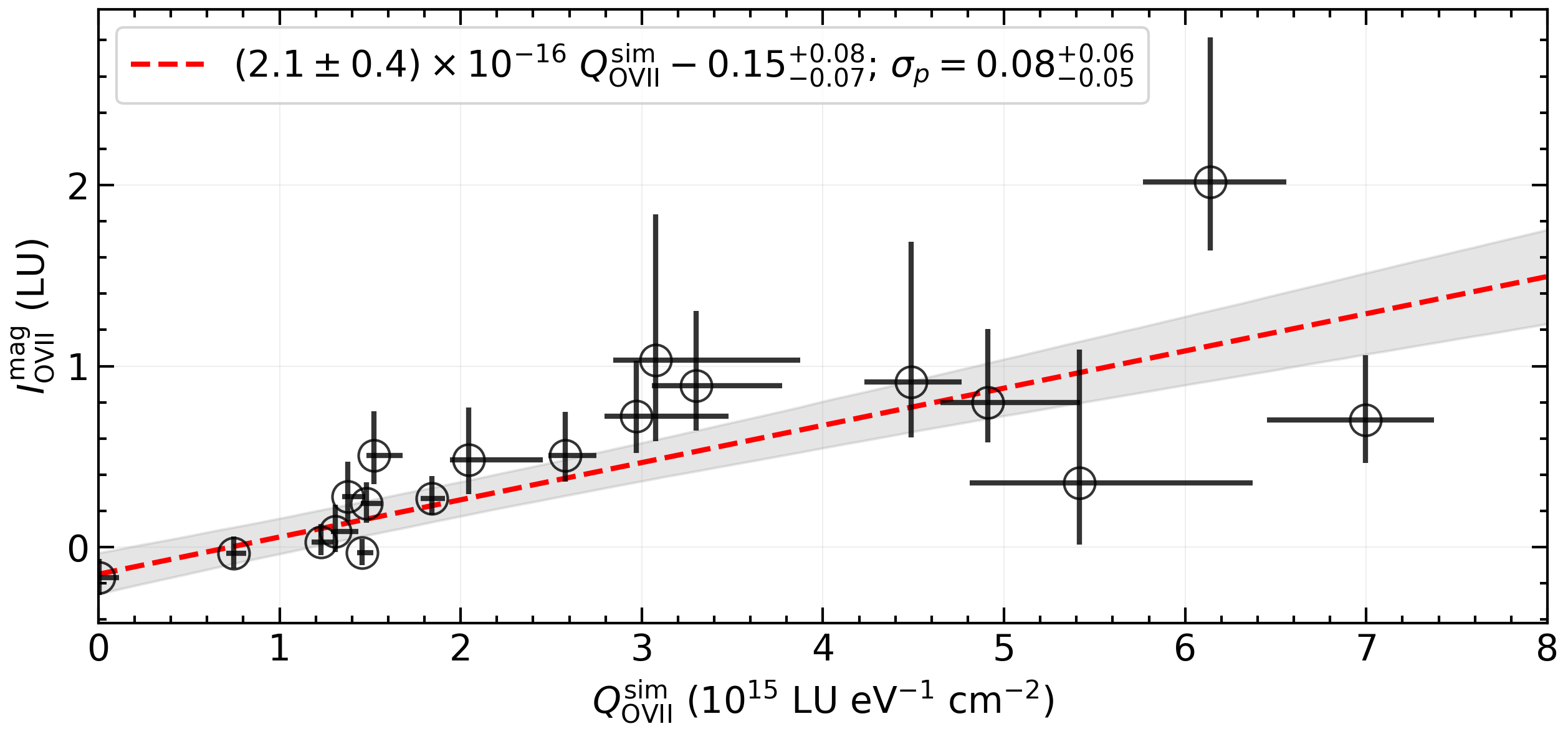}
\caption{The corrected \ion{O}{7} intensity correlates strongly with simulated total SWCX intensity, indicating that the simulation accurately reproduces the LOS integral of $n_{\rm H} n_{\rm SW} v_{\rm SW}$. This correlation suggests that the \ion{O}{7} line contributes approximately 20\% of total SWCX emission. The best-fit slope yields an empirical efficiency of $\alpha_{\rm OVII} \approx (2.1 \pm 0.4) \times 10^{-16}~{\rm eV~cm^2}$.
All data points are derived from the Voronoi bins shown in Figure~\ref{fig:spatial}.}

\label{fig:alpha}
\end{figure*}
\section{Discussion and Summary}
\label{sec:Discussion and Summary}
Using \xmm observations over 22 years, we systematically investigate the magnetospheric SWCX emission through the soft X-ray \ion{O}{7} line.
We present the spatial distribution of the intensity measured at the satellite positions (Figure~\ref{fig:spatial}, left), which exhibits a prominent enhancement at $X_{\rm GSE} \approx 10 \ \rm R_E$. This location is consistent with the expected subsolar magnetosheath region \citep[e.g.,][]{Fuselier2010Geo}, and the overall spatial pattern agrees closely with simulated magnetospheric SWCX signals (Figure~\ref{fig:spatial}, right), confirming its magnetospheric origin.

The observed emission also demonstrates a clear directional dependence (Figure \ref{fig:Angle}), with longer path lengths through the magnetosheath resulting in stronger signals. The most intense emissions, reaching up to $\approx$2.0 LU, occur when sightlines pass through the subsolar region of the magnetosheath.
Leveraging these dependencies, we construct an empirical magnetosheath model with boundaries fitted using the functional form from \citet{Shue1997} and emissivity fitted using a Gaussian function (Figure \ref{fig5:2d3dshape}).

Quantifying the magnetospheric emission on the nightside remains challenging, as the expected signal drops rapidly anti-sunward. We estimate the threshold beyond which this emission becomes negligible using $\phi$, the angle between the sunward ($+X_{\rm GSE}$) direction and the vector from Earth to the LOS magnetopause intersection ($\phi = 0^\circ$ corresponds to LOS pointing to the subsolar point, $\phi = 180^\circ$ anti-sunward). 
Specifically, $\phi$ values for observations inside the magnetopause are calculated based on the fitted magnetopause geometry ($r_0^{\rm MP}=9.7 \ \rm R_E$ and $a^{\rm MP}=0.5$), and binned with bin width $=5^\circ$. The result shows that the median intensity drops from $\sim$0.4 LU (2$\sigma$ significance) to zero at $\phi\sim100^{\circ}$, suggesting magnetospheric emission becomes negligible for $\phi \gtrsim 100^{\circ}$. This result provides empirical support for the observing strategies adopted by X-ray missions to minimize magnetospheric SWCX contamination. For instance, HaloSat restricts its observations to $\phi>110^\circ$ \citep{halosat2019ApJ...884..162K} to mitigate potential magnetospheric contamination. However, this cutoff angle should be interpreted with caution, as the mean magnetopause geometry used here does not account for the real‑time shifts driven by SW dynamic pressure, introducing uncertainties in the estimated $\phi$.

We further derive an empirical charge-exchange efficiency for the \ion{O}{7} line ($\alpha_{\rm OVII}$) from the correlation between observed and simulated intensities. 
Figure~\ref{fig:alpha} compares $I^{\rm mag}_{\rm OVII}$ and $Q^{\rm sim}_{\rm OVII}$ across all Voronoi bins defined in Figure~\ref{fig:spatial}, with each point representing the median and $1\sigma$ uncertainty.
The two quantities exhibit a clear linear relation, indicating that the true LOS integral of the interaction rate $Q$ (introduced in Section \ref{subsec:simulation}) is well reproduced by the simulation $Q^{\rm sim}_{\rm OVII}$, leading to $I^{\rm mag}_{\rm OVII} \approx \alpha_{\rm OVII} Q^{\rm sim}_{\rm OVII}$ according to Equation~\ref{eq:SWCX_I}. Therefore, the parameter $\alpha_{\rm OVII}$ can be empirically estimated by fitting this relationship with a linear model:
\begin{equation}
I^{\rm mag}_{\rm OVII} = \alpha_{\rm OVII} Q^{\rm sim}_{\rm OVII} + b
\end{equation}

The intercept $b$ accounts for a potential systematic offset between the data and the simulation due to our limited understanding of the absolute zero level of this emission. This fitting employs the likelihood presented in Section 2.4 of \citet{Sharma2017ARA&A}, which accounts for uncertainties in both axes and captures intrinsic scatter in the data using a patchiness parameter $\sigma_p$. 
The best-fit parameters are $\alpha_{\rm OVII} =(2.1\pm0.4)\times10^{-16}~\rm eV~cm^2$, $b = -0.15^{+0.08}_{-0.07}~\rm LU$, and $\sigma_p = 0.08^{+0.06}_{-0.05}~\rm LU$. The small but non-zero intercept $b$ indicates a minor zero-point offset in the measured magnetospheric SWCX. Accounting for it in the fit yields a more reliable estimate of the slope.

The derived $\alpha_{\rm OVII}$ provides an empirical constraint for the absolute intensity scale of magnetospheric SWCX models.
It is about one fifth of the scaling factor assumed in the MHD simulation, as the latter represents the total soft X-ray SWCX efficiency including contributions from multiple ions and emission lines. This ratio suggests that the \ion{O}{7} triplet accounts for $\sim$20\% of the total soft X-ray SWCX emission efficiency, rather than indicating an anomalously weak magnetospheric SWCX signal.
The derived value is broadly consistent with previous theoretical estimates based on ACE measurements, despite differences in the adopted atomic data. 
Specifically, it slightly exceeds the overall average $\sim 1 \times 10^{-16} \rm \ eV \ cm^2$ \citep{Liang2025},  and falls within the uncertainty of $\alpha$ in the 0.5--0.7 keV band for the streamer wind ($1.2^{+1.9}_{-0.8} \times 10^{-16} \rm \ eV \ cm^2$; \citealt{Koutroumpa2024E}). The value is notably larger than the estimate for coronal hole (CH) wind ($0.3^{+0.3}_{-0.1}\times 10^{-16}\rm \ eV \ cm^2$; \citealt{Koutroumpa2024E}), which occurs more often around solar maximum. This discrepancy is expected, as streamer wind carries a higher $O^{+7}/O^{+6}$ ratio than CH wind, and is the dominant wind type during \xmm observations.
To confirm this, we identify the SW type that reached the magnetosphere during each observation. Specifically, we use the SW proton speed as a proxy for wind type, since slow and fast SW broadly correspond to streamer-belt and CH origins, respectively, although a more rigorous classification would require charge-state composition \citep{von2016ApJ...816...13V}.
We match each observation midpoint to the nearest ACE/SWICS speed measurement at L1 \citep{Gloeckler1998SSRv...86..497G}, propagated to Earth.
Among the matched observations, $\sim87\%$ were taken under slow/streamer-like wind conditions ($v_{\rm SW}<550~{\rm km~s^{-1}}$), whereas only $\sim2\%$ were taken under fast/CH-like wind conditions ($v_{\rm SW}>650~{\rm km~s^{-1}}$).
This distribution confirms that the \xmm sample is dominated by slow-wind conditions over the solar cycle, weighting the empirical $\alpha_{\rm OVII}$ toward the streamer-wind value.

We note that the derived $\alpha_{\rm OVII}$ depends on the accuracy of the modeled $Q^{\rm sim}_{\rm OVII}$, which in turn is sensitive to the adopted exospheric neutral hydrogen density profile, $n_{\rm H}=25~{\rm cm^{-3}}(10~{\rm R_E}/r)^3$ \citep{Cravens2000ApJ}. The profile declines with $r^{-3}$ and is broadly consistent with Lyman-$\alpha$ observations \citep{Zoennchen2011AnGeo, Baliukin2019JGRA..124..861B}. 
However, the absolute density remains uncertain, especially beyond $\sim8~{\rm R_E}$, where geocoronal Lyman-$\alpha$ measurements become increasingly contaminated by the interplanetary background \citep{Bailey2011J}. For instance, the estimated absolute density at 10 $\rm R_E$ can range from $\sim4$ to $18~{\rm cm^{-3}}$ \citep{Zoennchen2013A}.
Thus, the derived $\alpha_{\rm OVII}$ should be interpreted as a model-dependent result that uses the specific neutral density model.

The upcoming SMILE mission (\citealt{SMILE2025SSRv..221....9W, SMILE2025SSRv..221...46C}) is specifically designed to study dynamic SW–magnetosphere interactions using magnetospheric SWCX.
Its Soft X-ray Imager (SXI) has a wide FOV ($15.6^\circ \times 26.5^\circ$) and good spatial resolution (FWHM$=6'$), allowing it to trace dynamic interactions between various SW ions and geocoronal neutrals in real time. These observations will place tighter constraints on the spatial distributions of both populations in near-Earth space during the SW-magnetosphere interaction, and its impact on space weather.

\begin{acknowledgments}
We are grateful to the anonymous referee for the constructive and insightful comments that significantly improved this work.
We thank Guiyun Liang for valuable insights regarding charge exchange efficiency and Dimitra Koutroumpa for valuable comments on this work.
ZP and JL acknowledge support from the NSFC through grant Nos. 12588202.
ZQ and LJ acknowledge the support of the National Key R\&D Program of China 2025YFF0512100.
TS and YG acknowledge support from the NSFC through grant Nos. 42322408 and 42188101.
JNB would like to acknowledge the University of Michigan for its support.
JL acknowledges support from the New Cornerstone Science Foundation through the New Cornerstone Investigator Program and the XPLORER PRIZE.
This research is based on observations obtained with XMM-Newton, an ESA science mission with instruments and contributions directly funded by ESA Member States and NASA.
\end{acknowledgments}

\bibliography{sample7}{}
\bibliographystyle{aasjournalv7}

\end{document}